\documentclass[conference]{IEEEtran}
\IEEEoverridecommandlockouts

\usepackage{cite}
\usepackage{amsmath,amssymb}
\usepackage{graphicx}
\usepackage{booktabs}
\usepackage{url}
\usepackage[hidelinks]{hyperref}
\hypersetup{
  pdftitle={Process Mining of Patient Flow in an Emergency Department: Bottlenecks, Variant Fragmentation, and Data-Driven Redesign},
  pdfauthor={Iliano Fasolino},
  pdfsubject={Process mining analysis of emergency department patient flow},
  pdfkeywords={process mining, emergency department, event log, conformance checking, healthcare processes, bottleneck analysis, patient flow}
}
\usepackage{tikz}
\usetikzlibrary{arrows.meta,positioning}

\graphicspath{{figures/}}

\begin{document}

\title{Process Mining of Patient Flow in an Emergency Department:\\ Bottlenecks, Variant Fragmentation, and Data-Driven Redesign}

\author{\IEEEauthorblockN{Iliano Fasolino}
\IEEEauthorblockA{Department of Computer Science\\
University of Milan\\
Milan, Italy}}

\maketitle

\begin{abstract}
Emergency departments (EDs) run some of the least standardized processes in healthcare, and long stays are often attributed to demand rather than measured on data. We analyze an anonymised event log of 1{,}820 ED stays (25{,}115 events) through a full process mining pipeline: preprocessing resolves burst logging and missing values, performance analysis quantifies flow at the case and transition level, and Inductive Miner with token-based conformance checking assesses process structure. The filtered log (1{,}754 cases, 16{,}376 events) exhibits a mean throughput of 6.58 hours with a heavy tail, 884 distinct control-flow variants whose most frequent one covers only 4.3\% of cases, and an inversion of clinical priority in which urgent patients (acuity 2, 8.03 h) stay 38\% longer than critical ones (acuity 1, 5.83 h). Perfect fitness (1.0) combined with low precision (0.71) reveals absence of normative pathways rather than deviance. Two redesign scenarios, acuity-based care pathways and anticipatory discharge, are derived with quantified targets.
\end{abstract}

\begin{IEEEkeywords}
process mining, emergency department, event log, conformance checking, healthcare processes, bottleneck analysis
\end{IEEEkeywords}

\section{Introduction}
\label{sec:intro}

Crowding in emergency departments is a structural problem with documented effects on mortality, treatment delays, and patient satisfaction \cite{hoot2008}. The patient journey in an ED spans arrival, triage, clinical evaluation, repeated monitoring, medication handling, and discharge or admission. Each of these stages leaves a digital footprint in hospital information systems, which makes the ED a natural target for process mining, the discipline that extracts process knowledge from event logs \cite{aalst2016}.

Process mining has been applied to healthcare for over a decade. Rojas et al. \cite{rojas2016} reviewed 74 case studies and observed that control-flow discovery dominates the literature, while conformance checking and redesign receive far less attention. Mans et al. \cite{mans2008} mined the gynaecological oncology process of a Dutch hospital and reported the "spaghetti" structure typical of clinical logs. Partington et al. \cite{partington2015} compared chest-pain pathways across four Australian hospitals and showed that cross-site variation is measurable from routinely collected data. These studies share a recurring outcome: hospital processes are far less structured than administrative ones, and discovered models alone rarely translate into operational change.

The gap we address is the connection between diagnosis and intervention. Discovering that a clinical process is fragmented is not sufficient; a redesign proposal needs quantified evidence of where time is lost, which patient segment suffers most, and what target values an intervention should reach. We work on an anonymised event log of 1{,}820 patient stays in an ED and carry the analysis through four stages: preprocessing, performance analysis, discovery with conformance checking, and scenario design.

The contributions are the following. First, a preprocessing pipeline that treats burst logging (multiple events with identical timestamps) as a modeling decision rather than a data quality error, reducing the log by 33\% while preserving the macro flow (Section~\ref{sec:method}). Second, a performance analysis that exposes a degenerate KPI (time-to-triage constant at one second, a logging artifact), an administrative discharge delay of 48 minutes on average, and a priority inversion between acuity levels 1 and 2 (Section~\ref{sec:results}). Third, a conformance argument: fitness 1.0 with precision 0.71 on an Inductive Miner model demonstrates that the ED lacks a normative process definition, since the model can only describe behavior, not constrain it. Fourth, two redesign scenarios with measurable targets derived from the mined evidence (Section~\ref{sec:redesign}).

\section{Method}
\label{sec:method}

\subsection{Event Log and Preprocessing}
\label{sec:preprocessing}

The dataset is an anonymised event log of ED stays provided for academic use. A case is a single patient stay identified by \texttt{stay\_id}; an event is the execution of an activity at a recorded timestamp. Six activity labels occur in the log: \emph{Enter the ED}, \emph{Triage in the ED}, \emph{Vital sign check}, \emph{Medicine dispensations}, \emph{Medicine reconciliation}, and \emph{Discharge from the ED}. The raw log contains 25{,}115 events over 1{,}820 cases. Case-level attributes (gender, race, arrival transport, acuity assigned at triage on a 1--5 scale, chief complaint, disposition) remain constant within a stay, whereas event-level attributes (vital signs, drug information) apply only to specific activities.

\begin{figure}[!t]
\centering
\begin{tikzpicture}[
  node distance=3.5mm,
  stage/.style={draw, rounded corners=1pt, fill=blue!6, text width=0.72\linewidth,
                align=center, font=\footnotesize, inner sep=3pt},
  note/.style={font=\scriptsize\itshape, text width=0.24\linewidth, align=left},
  arr/.style={-{Stealth[length=2mm]}, thick}
]
\node[stage] (raw) {Raw event log\\ \textbf{25{,}115 events, 1{,}820 cases}};
\node[stage, below=of raw] (sel) {Attribute selection\\ 9 columns retained (case ID, timestamp, activity + 6 case-level attributes)};
\node[stage, below=of sel] (agg) {Aggregation of simultaneous events\\ \textbf{16{,}826 events} ($-$33\%)};
\node[stage, below=of agg] (fill) {Case-level attribute completion\\ forward/backward fill: missing 89\% $\rightarrow$ 0\%};
\node[stage, below=of fill] (struct) {Structural filter\\ complete start/end, $\geq 4$ events\\ \textbf{1{,}771 cases} (49 removed)};
\node[stage, below=of struct] (dur) {Duration filter (0.5\,h--48\,h)\\ \textbf{1{,}754 cases, 16{,}376 events} (17 removed)};
\draw[arr] (raw) -- (sel);
\draw[arr] (sel) -- (agg);
\draw[arr] (agg) -- (fill);
\draw[arr] (fill) -- (struct);
\draw[arr] (struct) -- (dur);
\end{tikzpicture}
\caption{Preprocessing pipeline with exact event and case counts after each stage. In total 66 cases (3.6\%) are discarded as noise; the aggregation step accounts for the entire 33\% event reduction and no clinical activity type is lost.}
\label{fig:pipeline}
\end{figure}
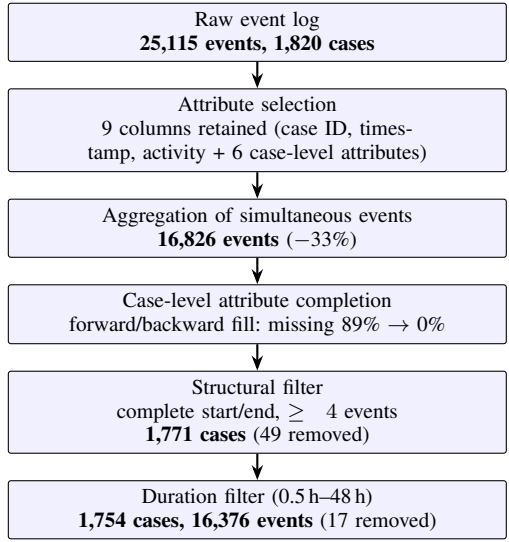

Three preprocessing decisions shape everything downstream, so we justify each one explicitly. Figure~\ref{fig:pipeline} summarizes the pipeline with the exact counts at every stage.

\emph{Simultaneous events.} Many stays contain two or more events with the same timestamp, typically medication records. We do not treat them as logging errors: administering several drugs at once is normal clinical practice. Since the analytical goal is the end-to-end macro flow (enter, triage, treatment, discharge) and not the pharmacological detail of each administration, we aggregate identical-timestamp events of the same case into one event. The choice trades granularity for focus: the log shrinks from 25{,}115 to 16{,}826 events (33\%) while every activity type and every temporal transition survives. The alternative, keeping the bursts, would inflate self-loop frequencies in discovery without adding control-flow information.

\emph{Missing values.} Around 89\% of the attribute cells are empty. Inspection shows a systematic pattern rather than random loss: acuity, gender, and race are recorded only at triage; vital signs only during \emph{Vital sign check}; drug fields only during medication activities. We therefore distinguish case-level from event-level attributes. Case-level attributes are propagated within each case with forward fill followed by backward fill, which brings their missing rate from 89\% to 0\%; event-level attributes keep their NaN values because imputing a heart rate onto a triage event would be meaningless.

\emph{Noise filtering.} Three rules define noise. Cases must contain \emph{Enter the ED} and end with \emph{Discharge from the ED}; all 1{,}820 cases pass this check, which tells us the logging of case boundaries is reliable. Cases with fewer than four events cannot represent a complete enter-triage-treatment-discharge path, so 49 such cases are removed as probable logging errors. Finally, durations below 30 minutes are unrealistic for a complete ED visit, and durations above 48 hours correspond to prolonged boarding, an exceptional regime that would distort the statistics of the standard flow; 6 cases fall below and 11 above these bounds. The filtered log holds 1{,}754 cases and 16{,}376 events.

\subsection{Performance Metrics}
\label{sec:metrics}

For a case $c$ with ordered events $e_1, \dots, e_m$ at timestamps $t_1 \leq \dots \leq t_m$, throughput time is defined as
\begin{equation}
T(c) = t_m - t_1 .
\label{eq:throughput}
\end{equation}
Time-to-discharge isolates the administrative tail of the stay: with $t_D$ the timestamp of \emph{Discharge from the ED} and $t_{D-1}$ the timestamp of the last activity preceding it,
\begin{equation}
\Delta(c) = t_D - t_{D-1} .
\label{eq:ttd}
\end{equation}
The same construction applied to \emph{Enter the ED} and \emph{Triage in the ED} yields time-to-triage. For bottleneck analysis we compute, for every pair of consecutive activities $(a, b)$ within a case, the waiting time of the transition, and aggregate over all occurrences:
\begin{equation}
W(a \rightarrow b) = \frac{1}{|O_{ab}|} \sum_{(e_i, e_{i+1}) \in O_{ab}} \left( t_{i+1} - t_i \right),
\label{eq:transition}
\end{equation}
where $O_{ab}$ is the set of occurrences of the transition in the log. We report means together with medians and occurrence counts, because waiting times in the log are strongly right-skewed and a mean alone would overstate typical behavior.

A variant is the ordered sequence of activity labels of a case. Variant analysis tests whether the process follows the Pareto expectation of well-structured processes, in which roughly 10--20\% of the variants cover 80\% of the cases; a strong violation of this expectation signals fragmentation rather than controlled flexibility.

\subsection{Process Discovery}
\label{sec:discovery}

We discover the control-flow model with the Inductive Miner \cite{leemans2013}, which recursively splits the directly-follows graph and guarantees sound, block-structured models. Robustness to infrequent behavior is the deciding property here: with 884 variants (Section~\ref{sec:variants}), algorithms without soundness guarantees produce unreadable spaghetti models on this log. The discovered process tree is converted to a Petri net for conformance analysis. All mining steps use PM4Py \cite{berti2019}; complementary map-style visualizations were produced with Disco, whose fuzzy-mining approach \cite{gunther2007} supports interactive filtering and animation for exploratory inspection.

\subsection{Conformance Checking}
\label{sec:conformance}

Conformance is evaluated with token-based replay \cite{rozinat2008}. Replaying a trace on the Petri net counts produced ($p$), consumed ($c$), missing ($m$), and remaining ($r$) tokens, and fitness is
\begin{equation}
f = \frac{1}{2}\left(1 - \frac{m}{c}\right) + \frac{1}{2}\left(1 - \frac{r}{p}\right).
\label{eq:fitness}
\end{equation}
Fitness alone is not informative for permissive models, so we complement it with precision in the ETConformance style \cite{munoz2010}, which penalizes model behavior never observed in the log, and with generalization and simplicity, following the four-dimension quality framework of Buijs et al. \cite{buijs2012}. The combination matters: a model can reach perfect fitness simply by allowing too much, and only the precision score exposes this.

\section{Experiments and Results}
\label{sec:results}

\subsection{Dataset After Preprocessing}

Table~\ref{tab:preprocessing} reports the effect of each preprocessing stage. The 3.6\% case loss is small and traceable: no case failed the boundary check, 49 were too short to be complete visits, and 17 fell outside the plausible duration window. Everything that follows uses the filtered log of 1{,}754 cases.

\begin{table}[!t]
\caption{Effect of each preprocessing stage on the event log.}
\label{tab:preprocessing}
\centering
\begin{tabular}{lrr}
\toprule
Stage & Events & Cases \\
\midrule
Raw log & 25{,}115 & 1{,}820 \\
Aggregation of simultaneous events & 16{,}826 & 1{,}820 \\
Structural filter ($\geq 4$ events, boundaries) & 16{,}679 & 1{,}771 \\
Duration filter (0.5\,h--48\,h) & 16{,}376 & 1{,}754 \\
\midrule
Total reduction & $-34.8\%$ & $-3.6\%$ \\
\bottomrule
\end{tabular}
\end{table}

\subsection{Design Choices}

\begin{figure*}[!t]
\centering
\includegraphics[width=\textwidth]{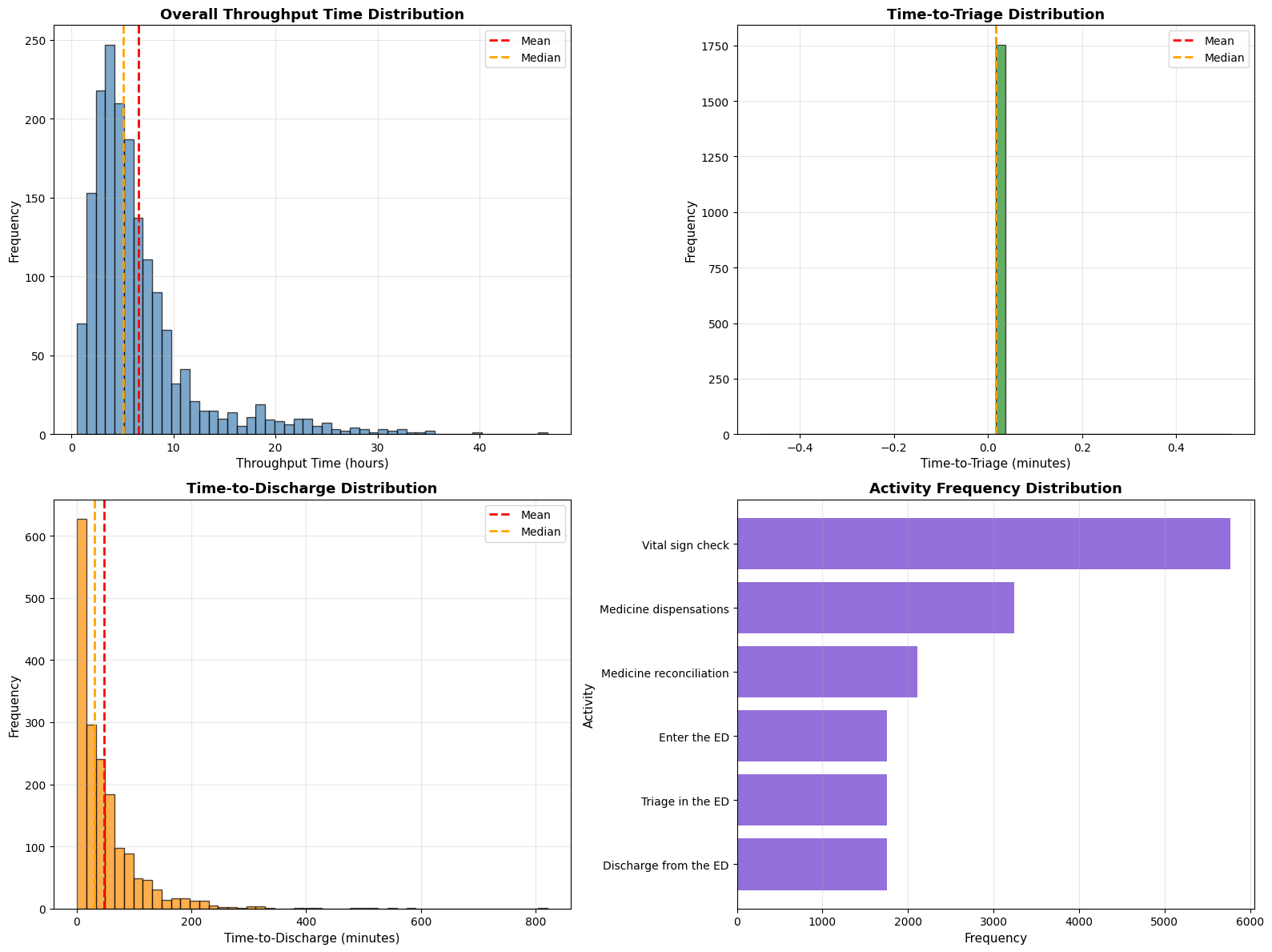}
\caption{Performance overview of the filtered log (1{,}754 cases). Top left: throughput time distribution with mean 6.58 h (red) and median 5.08 h (orange); the right tail extends to 46.6 h. Top right: the degenerate time-to-triage distribution, constant at one second for every case, which identifies a logging artifact rather than a process measurement. Bottom left: time-to-discharge distribution with mean 48.3 min and maximum 820.7 min. Bottom right: activity frequencies; \emph{Vital sign check} dominates with 5{,}764 occurrences, more than three times the case count.}
\label{fig:overview}
\end{figure*}

Two decisions deserve emphasis before the numbers. The duration window (0.5--48 h) was set from domain reasoning, not from percentile clipping: a complete ED visit under 30 minutes is physically implausible, and stays beyond 48 hours belong to the boarding regime, which is a real phenomenon but a different process. Second, one planned KPI was dropped during analysis. Time-to-triage turned out to be exactly one second for all 1{,}754 cases with zero variance (Figure~\ref{fig:overview}, top right). No queue behaves this way; the only consistent explanation is that \emph{Enter the ED} and \emph{Triage in the ED} are written by the information system as a combined record. We classify time-to-triage as a logging artifact, exclude it from bottleneck analysis, and flag it as a data quality finding in its own right: a hospital that wants to monitor door-to-triage performance cannot do so with this system configuration.

\subsection{Temporal Performance}

Table~\ref{tab:kpi} summarizes the two reliable temporal KPIs. Mean throughput, computed with Eq.~\eqref{eq:throughput}, is 6.58 hours against a median of 5.08, with standard deviation 5.42 hours; 10\% of the patients leave within 2.2 hours while another 10\% stay beyond 12.4 hours. The distribution in Figure~\ref{fig:overview} (top left) shows the classic ED shape: a mode between 3 and 5 hours and a long boarding-driven tail.

Time-to-discharge, the interval between the last clinical activity and formal discharge, averages 48.3 minutes with a median of 31 and a maximum of 820.7 minutes, that is, one patient waited 13.7 hours after being clinically ready. In 79.1\% of the cases the last activity before discharge is a vital sign check, consistent with a stability confirmation preceding the discharge decision. The delay that follows is administrative by construction: paperwork, prescriptions, patient instructions, transport. This window adds no clinical value and affects every single case, which makes it a high-yield redesign target (Section~\ref{sec:scenario2}).

\begin{table}[!t]
\caption{Temporal KPIs on the filtered log. Time-to-triage is excluded as a logging artifact (constant 1\,s, $\sigma = 0$).}
\label{tab:kpi}
\centering
\begin{tabular}{lrr}
\toprule
Statistic & Throughput (h) & Time-to-discharge (min) \\
\midrule
Mean & 6.58 & 48.3 \\
Std.\ dev. & 5.42 & 61.8 \\
Minimum & 0.55 & 0.8 \\
25th percentile & 3.30 & 9.0 \\
Median & 5.08 & 31.0 \\
75th percentile & 7.70 & 61.3 \\
Maximum & 46.63 & 820.7 \\
\bottomrule
\end{tabular}
\end{table}

\subsection{Variant Analysis}
\label{sec:variants}

The 1{,}754 cases produce 884 distinct variants. The most frequent one, the minimal path \emph{Enter} $\rightarrow$ \emph{Triage} $\rightarrow$ \emph{Vital sign check} $\rightarrow$ \emph{Discharge}, covers 75 cases (4.3\%). Table~\ref{tab:variants} lists the ten most frequent variants; together they cover 21.3\% of the cases. Reaching 80\% coverage requires 519 variants and 90\% requires 686.

\begin{figure}[!t]
\centering
\includegraphics[width=\linewidth]{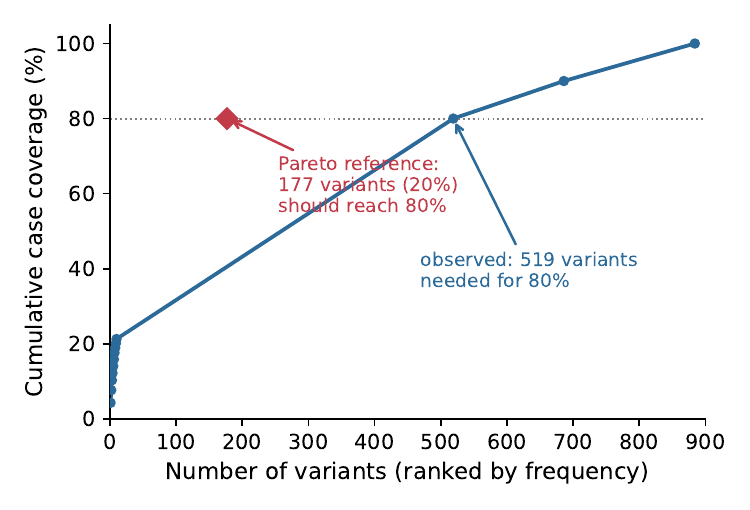}
\caption{Cumulative case coverage as a function of the number of variants, ranked by frequency. Markers are measured coverage points from the log (top-10 cumulative shares, then 519 variants at 80\%, 686 at 90\%, 884 at 100\%); segments interpolate between them. The red diamond marks the Pareto reference: a structured process would reach 80\% coverage with about 177 variants (20\%), while this log requires 519.}
\label{fig:coverage}
\end{figure}

\begin{figure*}[!t]
\centering
\includegraphics[width=0.95\textwidth]{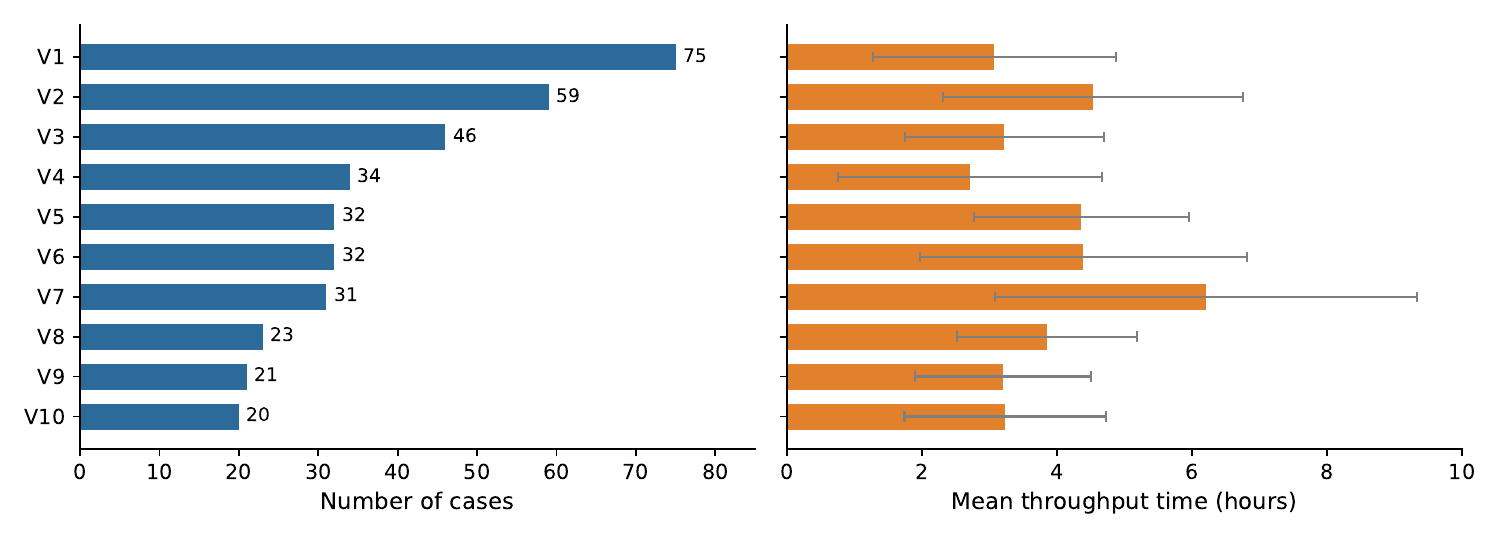}
\caption{The ten most frequent variants (labels as in Table~\ref{tab:variants}). Left: case frequency; even the most common path covers only 75 of 1{,}754 cases. Right: mean throughput time with one standard deviation; V7, three consecutive vital sign checks, is the slowest frequent variant at 6.21 h, which anticipates the monitoring loop identified as the primary bottleneck.}
\label{fig:variants}
\end{figure*}

Figure~\ref{fig:coverage} sets these numbers against the Pareto expectation. In a structured process, 20\% of the variants (177 here) would cover about 80\% of the cases; this log needs 2.9 times as many. The violation is severe and it changes the interpretation of everything else: the ED does not run a small set of pathways with occasional exceptions, it runs an almost per-patient process. Figure~\ref{fig:variants} adds the duration dimension for the top ten variants. Even among these frequent paths, mean durations range from 2.7 to 6.2 hours, and V7 (three consecutive vital sign checks) is the slowest at 6.21 hours, an early hint that repeated monitoring is where time accumulates.

\begin{table}[!t]
\caption{Ten most frequent variants. Activities are abbreviated as E = Enter the ED, T = Triage, V = Vital sign check, MD = Medicine dispensations, MR = Medicine reconciliation, D = Discharge.}
\label{tab:variants}
\centering
\begin{tabular}{llrr}
\toprule
ID & Sequence & Cases & Share \\
\midrule
V1 & E, T, V, D & 75 & 4.3\% \\
V2 & E, T, V, V, D & 59 & 3.4\% \\
V3 & E, T, MD, V, D & 46 & 2.6\% \\
V4 & E, T, MR, V, D & 34 & 1.9\% \\
V5 & E, T, V, MR, V, D & 32 & 1.8\% \\
V6 & E, T, V, MD, V, D & 32 & 1.8\% \\
V7 & E, T, V, V, V, D & 31 & 1.8\% \\
V8 & E, V, T, V, D & 23 & 1.3\% \\
V9 & E, T, MR, MD, V, D & 21 & 1.2\% \\
V10 & E, V, T, MD, V, D & 20 & 1.1\% \\
\midrule
Total top 10 & & 373 & 21.3\% \\
\bottomrule
\end{tabular}
\end{table}

\subsection{Segmentation by Acuity}
\label{sec:acuity}

\begin{figure}[!t]
\centering
\includegraphics[width=\linewidth]{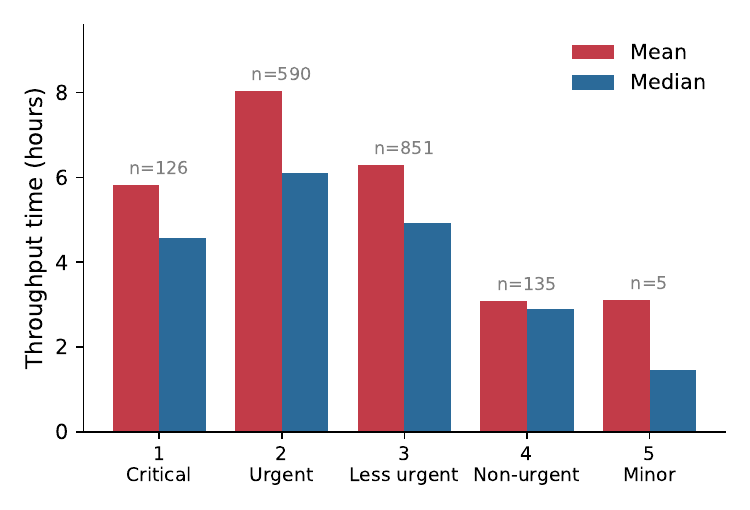}
\caption{Mean and median throughput time by acuity level with case counts. Acuity 2 (urgent) patients stay on average 8.03 h, exceeding acuity 1 (critical) patients at 5.83 h and inverting the clinically expected ordering; the gap persists in the medians (6.10 h against 4.58 h), so it is not driven by a few outliers.}
\label{fig:acuity}
\end{figure}

Segmenting throughput by the triage acuity score produces the most consequential finding of the performance phase. Figure~\ref{fig:acuity} shows mean and median throughput per acuity level. Critical patients (acuity 1, $n = 126$) average 5.83 hours. Urgent patients (acuity 2, $n = 590$) average 8.03 hours, 38\% more, and their maximum reaches 46.6 hours. The ordering expected from clinical priority is inverted between the two highest urgency classes, while lower-acuity patients behave as expected (acuity 4, $n=135$: 3.08 h).

The inversion is not explained by case volume alone. A plausible mechanism, consistent with the bottleneck evidence below, is a prioritization gap: acuity 1 patients receive immediate, continuously attended care, and acuity 4--5 patients follow short paths, while acuity 2 patients are started on diagnostic workups and then repeatedly parked whenever critical arrivals preempt staff. Diagnostic complexity and boarding while awaiting admission are contributing candidates. Whatever the mix of causes, 590 patients, one third of the volume, systematically wait longer than patients classified as more severe. This is a structural inefficiency, not noise.

\subsection{Bottleneck Identification}
\label{sec:bottlenecks}

\begin{figure}[!t]
\centering
\includegraphics[width=\linewidth]{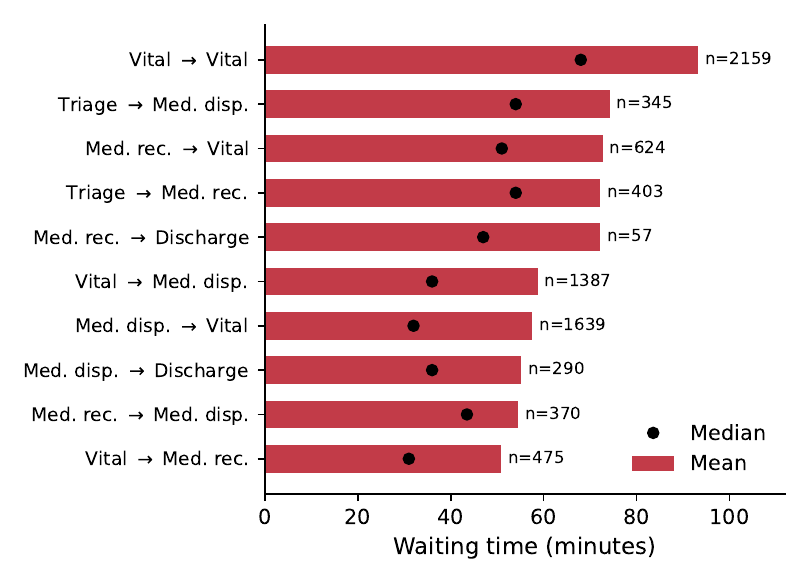}
\caption{Mean waiting time of the ten slowest frequent transitions ($\geq 50$ occurrences), with medians (dots) and occurrence counts. The vital-sign self-loop combines the longest mean wait (93.3 min) with the highest frequency (2{,}159 occurrences) and is the primary bottleneck; all distributions are right-skewed, hence medians sit well below means.}
\label{fig:bottlenecks}
\end{figure}

Applying Eq.~\eqref{eq:transition} to all consecutive activity pairs and restricting to transitions with at least 50 occurrences yields the ranking in Figure~\ref{fig:bottlenecks}. The self-loop \emph{Vital sign check} $\rightarrow$ \emph{Vital sign check} dominates on both axes: 93.3 minutes of mean waiting (median 68) over 2{,}159 occurrences, an order of magnitude more volume than most other transitions. These are patients in observation, waiting between repeated assessments. The pattern is compatible with two readings, insufficient nursing capacity for more frequent checks, or patients boarding in the ED while awaiting admission, consultation, or test results; the log does not carry resource data, so we cannot separate the two. Transitions out of triage into medication activities (74.2 and 72.2 minutes) form the second tier, and \emph{Vital sign check} $\rightarrow$ \emph{Discharge}, the administrative tail already quantified by Eq.~\eqref{eq:ttd}, occurs 1{,}387 times with a mean of 44.5 minutes. Weighted by frequency, the monitoring loop alone accounts for roughly 3{,}360 patient-hours of waiting in the log, more than any other transition by a factor of two.

\subsection{Discovered Model and Conformance}
\label{sec:model}

\begin{figure*}[!t]
\centering
\includegraphics[width=0.98\textwidth]{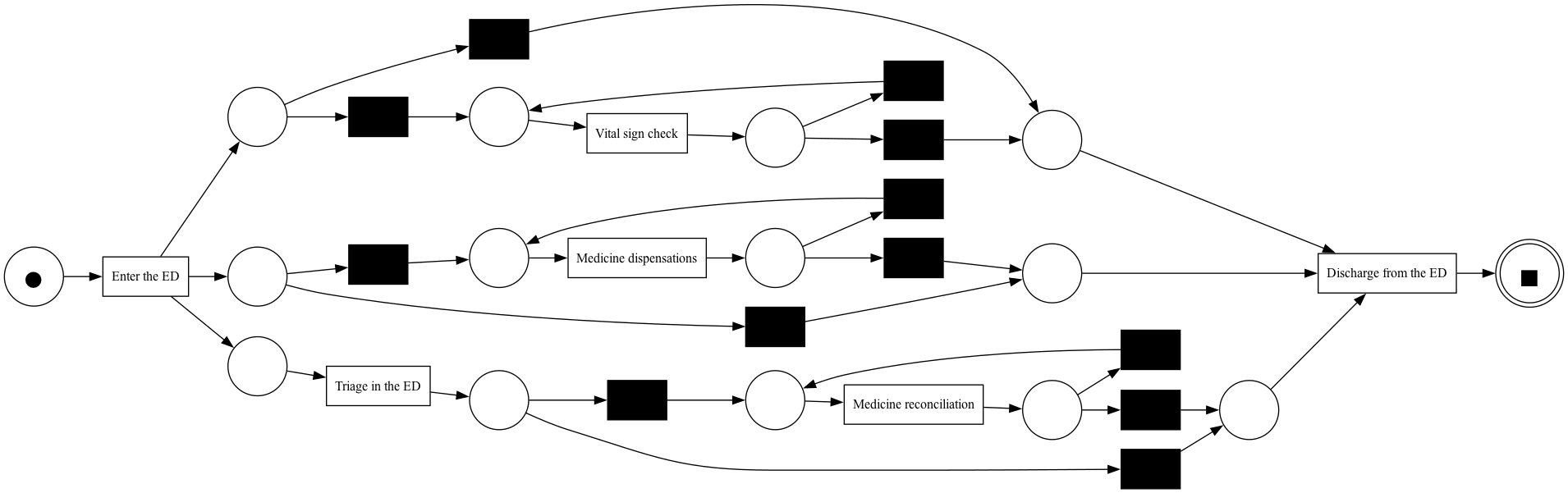}
\caption{Petri net discovered by the Inductive Miner (15 places, 18 transitions, 40 arcs; black rectangles are silent transitions). Three optional, loopable branches (vital sign monitoring, medicine dispensations, triage followed by medicine reconciliation) run between entry and discharge. The block structure guarantees soundness but admits many interleavings never observed in the log, which is what the precision score measures.}
\label{fig:petri}
\end{figure*}

\begin{figure*}[!t]
\centering
\includegraphics[width=0.9\textwidth]{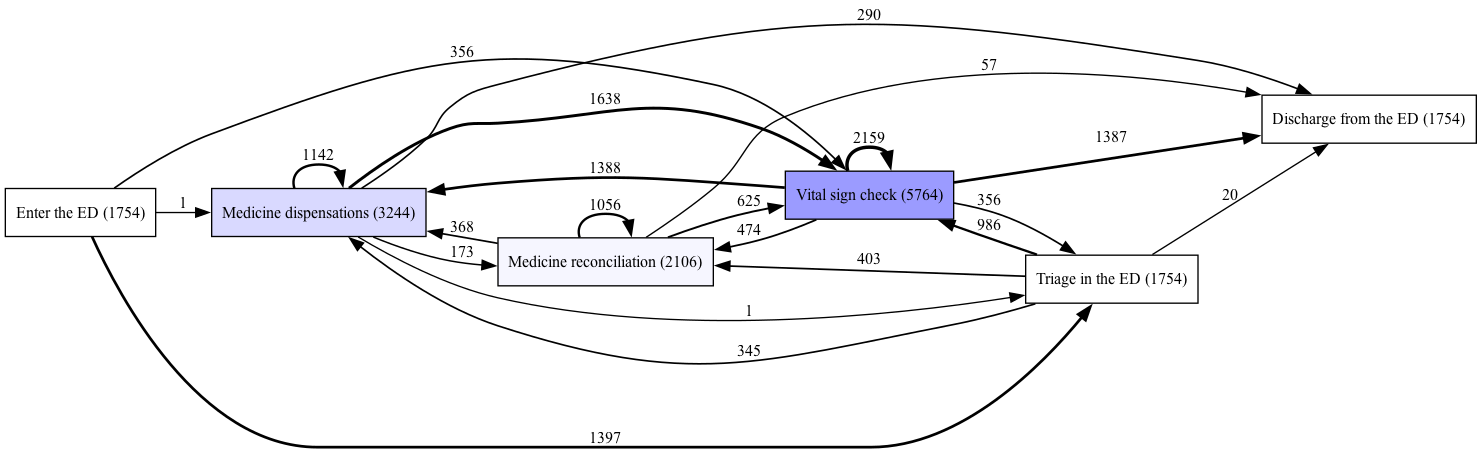}\\[2mm]
\includegraphics[width=0.9\textwidth]{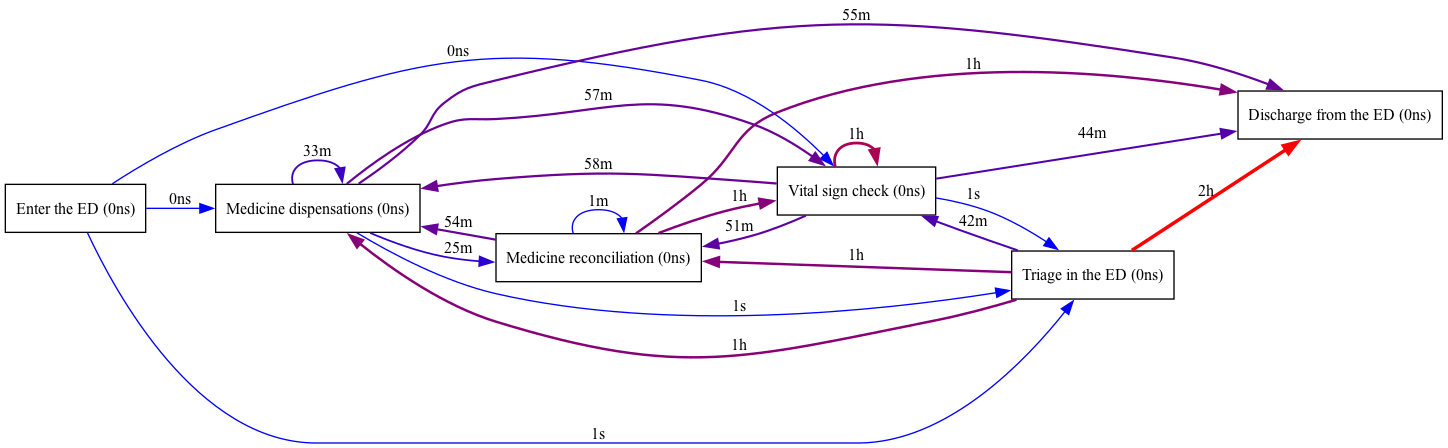}
\caption{Directly-follows graphs of the filtered log. Top: frequency view; \emph{Vital sign check} concentrates 5{,}764 events and a self-loop of 2{,}159, and 1{,}387 cases move directly from a vital sign check to discharge. Bottom: performance view (mean waiting per arc); the slowest arcs are the vital-sign self-loop (about 1 h) and the rare triage-to-discharge arc (2 h over 20 cases), matching the transition-level ranking of Figure~\ref{fig:bottlenecks}.}
\label{fig:dfg}
\end{figure*}

The Inductive Miner produces a process tree converted to a Petri net with 15 places, 18 transitions, and 40 arcs (Figure~\ref{fig:petri}). Its structure is readable: after entry, three concurrent branches cover triage, vital sign monitoring, and the two medication activities, each skippable and each with loop behavior, converging into a single discharge. The frequency and performance views of the directly-follows graph (Figure~\ref{fig:dfg}) quantify the same structure: the vital-sign node concentrates 5{,}764 occurrences with a self-loop of 2{,}159, and the slowest arcs coincide with the transitions of Figure~\ref{fig:bottlenecks}.

Token-based replay with Eq.~\eqref{eq:fitness} over all 1{,}754 cases yields the quality profile of Table~\ref{tab:conformance}: fitness 1.0000, precision 0.7139, generalization 0.9694, simplicity 0.7021, and zero deviant traces. Taken in isolation, perfect fitness sounds like good news. Combined with the variant statistics it means the opposite. A healthy, standardized process would show a handful of dominant variants, fitness around 0.90--0.95 against a normative model, precision above 0.85, and a small set of deviant cases worth investigating. Here the profile is inverted on every dimension: 884 variants, no deviants at all, and a model that must stay permissive (precision 0.71) to accommodate everything it saw. The model can describe what happens; nothing in the organization prescribes what should happen. Conformance checking therefore measures, quantitatively, the absence of a normative process definition, and this reframes the redesign question from enforcing an existing standard to creating one.

\begin{table}[!t]
\caption{Conformance and model quality metrics (token-based replay on 1{,}754 cases).}
\label{tab:conformance}
\centering
\begin{tabular}{lrl}
\toprule
Metric & Value & Reading \\
\midrule
Fitness & 1.000 & all traces replay without errors \\
Precision & 0.714 & model allows unseen behavior \\
Generalization & 0.969 & model not overfitted to the log \\
Simplicity & 0.702 & moderate structural complexity \\
Deviant traces & 0 / 1{,}754 & no normative baseline to deviate from \\
\bottomrule
\end{tabular}
\end{table}

\subsection{Comparison with Literature}

Our findings match the qualitative picture of clinical process mining while adding quantification that earlier reports lack. The fragmentation level (most frequent variant at 4.3\%) is of the same order as the gynaecological oncology log of Mans et al. \cite{mans2008}, where the dominant variant also covered a marginal share of cases. Partington et al. \cite{partington2015} found large inter-hospital variation in chest-pain management; we find comparable variation within a single department, across acuity strata. The reviewed literature \cite{rojas2016} reports that most healthcare studies stop at discovery; the conformance profile we report (perfect fitness, low precision, zero deviants) offers a reusable diagnostic signature for the "no normative model" condition that other EDs can test on their own logs. On the intervention side, our discharge targets are aligned with reductions of 50--60\% reported for anticipatory discharge planning in acute settings \cite{baker2012,leckcivilize2021}, and the fast-track separation we propose for low-acuity patients has documented effects on ED length of stay \cite{considine2008}.

\section{Redesign Scenarios}
\label{sec:redesign}

The evidence points at two distinct problems: fragmentation with a priority inversion (Sections \ref{sec:variants}--\ref{sec:acuity}) and an administrative discharge tail (Section~\ref{sec:bottlenecks}). We propose one scenario for each, with targets summarized in Table~\ref{tab:targets}.

\begin{table}[!t]
\caption{Quantified targets of the two redesign scenarios.}
\label{tab:targets}
\centering
\begin{tabular}{lrr}
\toprule
KPI & Current & Target \\
\midrule
Distinct variants & 884 & $<$ 50 \\
Coverage of top-10 pathways & 21.3\% & 80\% \\
Acuity 2 mean throughput & 8.03 h & 5.5 h \\
Acuity 3 mean throughput & 6.30 h & 4--5 h \\
Acuity 4--5 mean throughput & 3.08 h & $<$ 2 h \\
Model precision & 0.71 & $>$ 0.85 \\
Time-to-discharge (3 months) & 48 min & 30 min \\
Time-to-discharge (6 months) & 48 min & 20 min \\
\bottomrule
\end{tabular}
\end{table}

\subsection{Scenario 1: Acuity-Based Standard Care Pathways}
\label{sec:scenario1}

The proposal is to define four care pathways, one per acuity band, as frameworks specifying expected activities, target timeframes, and dedicated resources rather than rigid sequences. The goal is to compress 884 variants below 50 while keeping clinical flexibility, so that variability reflects patient condition instead of coordination gaps.

The acuity 1 pathway formalizes current practice, which already performs acceptably (5.83 h): immediate triage, no inter-step waits, monitoring every 15--30 minutes, and an admission decision within 90 minutes, with a 4--5 hour throughput target focused on variance reduction. The acuity 2 pathway is the priority of the intervention, because this class carries the inversion documented in Section~\ref{sec:acuity} and 34\% of the volume. It prescribes triage within 10 minutes, diagnostic ordering within 30, a diagnostic phase of 60--90 minutes, and a disposition decision within roughly 30 minutes of results, for a 5--6 hour target (a 25--31\% reduction). Two organizational elements support it: a dedicated urgent care team that is not preempted by critical arrivals, precisely the parking mechanism hypothesized for the inversion, and an early admission evaluation within 4 hours so that inpatient coordination starts before the workup completes. The acuity 3 pathway (48.5\% of volume) accepts an initial wait but bounds the path after start, targeting 4--5 hours through nursing protocols for minor conditions. The acuity 4--5 fast track separates a physically distinct lane run by physician assistants under standardized protocols, with a sub-2-hour target \cite{considine2008}; the volume share is 8\%, but the separation releases main-ED capacity for the classes above.

Implementation follows three phases: a 1--2 month pilot on the acuity 2 pathway and the fast track with weekly KPI monitoring, a 3--6 month extension to all levels including training and information system support, and a permanent monitoring phase in which process mining runs monthly on fresh logs. Success is verifiable directly on the mined KPIs: variants below 50, top-10 coverage at 80\%, acuity 2 mean below 5.5 hours, and precision above 0.85, the last one indicating that a normative model finally exists and describes practice tightly.

\subsection{Scenario 2: Discharge Process Optimization}
\label{sec:scenario2}

The second scenario attacks the 48-minute administrative tail with three low-to-medium cost interventions. Anticipatory discharge planning moves preparation from "patient ready" to "decision taken": once the physician decides that a patient will leave after a confirmation check, the nurse starts the discharge summary from pre-populated templates, the pharmacy is alerted, instruction materials are selected, and transport is booked, all during the final monitoring window. Analogous anticipatory interventions in acute care report reductions of 50--60\% in the targeted delay \cite{baker2012,leckcivilize2021}, and the required investment is limited to training and change management over 1--2 months. The digital discharge component removes manual steps: summaries auto-populated from the EHR, electronic prescriptions sent directly to the patient pharmacy, and instructions delivered via SMS or email, with a 3--6 month integration effort. Nurse-led discharge, for acuity 4--5 patients with a clear plan, delegates the final step to trained nurses under remote physician approval, freeing medical time for the complex cases of Scenario 1.

The targets are 30 minutes mean time-to-discharge after 3 months and 20 minutes after 6 (Table~\ref{tab:targets}). The direct effect on total throughput is modest, about 5\%, but it applies to 100\% of the cases, and the freed bed time converts into arrival capacity. Discharge is also the last interaction the patient remembers, so cutting a 48-minute idle wait to 20 minutes has a reputational effect beyond the KPI.

\section{Conclusion}
\label{sec:conclusion}

We analyzed 1{,}754 emergency department stays end to end with process mining and turned an unstructured event log into a quantified diagnosis: throughput of 6.58 hours with a heavy tail, a degenerate time-to-triage KPI exposing a logging artifact, 884 variants with the most frequent at 4.3\%, an inversion in which urgent patients wait 38\% longer than critical ones, a monitoring self-loop absorbing 93 minutes per occurrence over 2{,}159 occurrences, and a conformance profile (fitness 1.0, precision 0.71, zero deviants) that measures the absence of a normative process. Each redesign target in Table~\ref{tab:targets} is tied to one of these measurements, which is the main methodological point of the work: proposals inherit the credibility of the numbers they are derived from.

Several limitations bound the conclusions. The log covers a single department without resource or staffing data, so the two candidate explanations of the monitoring bottleneck (understaffing against boarding) cannot be separated. Aggregating simultaneous events removes intra-burst detail that a medication-level analysis would need. The noise thresholds (four events, 0.5--48 h) are domain-motivated but heuristic, and excluded boarding cases deserve their own study. The improvement targets rest on published case studies rather than on-site simulation, and the acuity 5 stratum ($n=5$) is too small for inference.

Future work is concrete. A natural next step is pattern-based feature generation for variant analysis: encoding each case with a binary vector over clinically meaningful properties (age bands, abnormal vital signs, diagnosis groups) and mining variants in that space would connect control-flow fragmentation to patient characteristics, and could confirm or reject the parking hypothesis for acuity 2. We also plan a discrete-event simulation of the two scenarios calibrated on the mined transition times, so that the targets of Table~\ref{tab:targets} can be validated before any organizational change is attempted.

\end{document}